\documentclass[11pt,article]{amsart}
\usepackage{amssymb}
\usepackage{amsfonts}
\usepackage{amsbsy,url}
\usepackage{latexsym}
\usepackage{amssymb,latexsym,amsmath,amsthm}
\usepackage{color}
\theoremstyle{plain}

 \theoremstyle{remark} 

\newcommand{\C}{{\mathbb C}}
\newcommand{\N}{{\mathbb N}}

\renewcommand{\vec}[1]{{\bf #1}}

\newtheorem {theo} {\bf Theorem} [section]
\newtheorem {prop} [theo] {\bf Proposition}
\newtheorem {coro} [theo] {\bf Corollary}
\newtheorem {lem} [theo] {\bf Lemma}

\newtheorem{exam} {\bf Example}[section]

\newcommand{\R}{\mathbb R\,}

\numberwithin{equation}{section}
\makeatletter
\@namedef{subjclassname@2020}{%
  \textup{2020} Mathematics Subject Classification}
\makeatother
\begin{document}
\title{Zero-Error Correctibility and Phase Retrievability for  Twirling Channels }

\author{Deguang Han}
\address{Department of Mathematics\\
University of Central Florida\\ Orlando, FL 32816}
\email{deguang.han@ucf.edu}
\author{Kai Liu}
\address{Department of Mathematics\\ University of Central Florida\\Orlando, FL 32816}
\email{kailiu@knights.ucf.edu}
\thanks{Deguang Han is partially supported by the NSF grant DMS-2105038.}

\date{\today}

%
\begin{abstract}  A twirling  channel is a quantum channel induced by a continuous unitary representation $\pi = \sum_{i}^{\oplus} m_i\pi_i$, where $\pi_i$ are inequivalent irreducible representations. Motivated by a recent work \cite{Twirling} on minimal mixed unitary rank of $\Phi_{\pi}$, we explore  the   connections  of the independence number, zero error capacity, quantum codes, orthogonality index and phase retrievability of the quantum channel $\Phi_{\pi}$ with the  irreducible  representation multiplicities  $m_i$, the  irreducible representation dimensions $\dim H_{\pi_i}$. In particular we show that the independence number of $\Phi_{\pi}$ is the sum of the multiplicities, the orthogonal index of $\Phi_{\pi}$ is exactly the sum of those representation dimensions, and the  zero-error capacity is equal to $\log (\sum_{i=1}^{d}m_i)$. We also present a lower bound for the phase retrievability in terms of the minimal length of phase retrievable frames for $\C^n$.
\end{abstract}
\maketitle

\section{Introduction}

A quantum channel $\Phi$  is  a  completely positive trace-preserving (CPTP) linear map from an operator system $B(H)$  to an operator system $B(K)$, which has a  Kraus representation of the form:
$$
\Phi(T) = \sum_{i=1}^{r}A_iTA_i^*, \ \forall T \in B(H)
$$
for some operators $A_1, . . . , A_r\in B(H, K)$. In this representation,  $A_1, ... , A_r$ are also referred as the Kraus operators of $\Phi$. For a quantum channel $\Phi$,  the Choi–Jamiołkowski matrix \cite{Choi-LAA1, Jami}  is the matrix  defined by
$$
C_{\Phi} = [ \Phi(E_{ij})]_{n\times n}
$$
where   $\{e_i\}_{i=1}^{n}$ is an orthonormal  basis of $H$ and $E_{ij}$ is  the rank-one operator $e_i\otimes e_j$. The  {\it Choi rank} of  $\Phi$mis the smallest integer $r$ from the Kraus representations which is equal to the rank of $C_{\Phi}$.

Covariant channels form a special and important type of quantum channels where certain symmetries are present in the quantum channel. In this paper we are interested in exploring the connections of some important concepts/quantities for a group representation induced quantum channels (also refers to twirling channels) with its irreducible  decomposition of the representation.

For a compact group $G$, a continuous function $\pi: G \to U(H)$ is called a (finite dimensional) unitary representation if  $\pi(gh) = \pi(g)\pi(h)$. A subspace $V$ of $H$  is called invariant if $\pi(g)x\in V$ for all $g\in G$ and $x\in V$. A representation $\pi$  is called irreducible if $0$ and $H$ are the only invariant subspaces. It is well-known that any unitary representation $\pi$  on a finite-dimensional Hilbert space $H$ is the direct sum of irreducible representations. More precisely, there exists a unitary operator $U$ on $H$ such that 
$$
U\pi(g)U^* = m_1\pi_1 \oplus ... \oplus m_d\pi_d
$$
where $m_i\in \Bbb{N}$ and  $\pi_1, ... , \pi_d$ are inequivalent irreducible unitary representations of $G$ acting on Hilbert spaces $H_1, ... , H_d$, respectively. Clearly we have $\dim H = \sum_{i=1}^{d}m_in_i$ where $n_i = \dim H_i$. With the help of the characterization of mixed unitary quantum channel by the complement channels it was proved recently in \cite{Twirling} that  a unitary representation $\pi$ induced quantum channel $\Phi_{\pi}$ has the minimal mixed unitary rank in the sense that its mixed unitary rank is the same as the Choi rank which is equal to 
 $r = \sum_{i=1}^{d} (\dim H_i)^2$.  Inspired by this, naturally one would like to know how the multiplicity vector $\vec{m}= (m_1, ... , m_d)$ and the dimensional vector $\vec{n} = (n_1, ... , n_d)$ of the representation are related to several  other concepts such as independence number, quantum codes, zero-error capability for the induced quantum channels. It is well-known  that independence numbers and quantum zero error capacity are  among the important quantities in quantum communication theory and they have been extensively studied in the literature c.f.\cite{BDSW, Kribs-0, Kribs-1,  Kribs-2, Paulsen}. Additionally we introduce and  explore some  ``dual versions" of these concepts that include the concepts of  orthogonality index and phase retrievability. The phase retrievability of a quantum channel $\Phi$, which was recently introduced in \cite{KaiLiu-2}, concerning the ability of distinguishing the pure states from the input system by a positive  operator valued measure (POVM) or observables from the output system, The main purpose of this note is to obtain precise characterizations for all the above mentioned quantities for twirling channels $\Phi_{\pi}$. More precisely we shall prove the following statements: 
 \begin{itemize}
 
 \item[(i)] $\alpha(\Phi_{\pi}) = \sum_{i=1}^{d} m_i$ is the independence number of $\Phi_{\pi}$, and the zero error capacity is equal to $\log (\sum_{i=1}^{d}m_i)$.
 
 
 \item[(ii)] $\beta(\Phi_{\pi}) = \max\{m_1, ... , m_d\}$ is  the largest number $m$ such that there exists a quantum code of dimension $m$.
 
 \item[(iii)] $\sum_{i=1}^{d}n_i$ is the orthogonality index of $\Phi_{\pi}$.
 
 \item[(iv)] $\max\{n_1, ... , d_d\}$  is  the orthogonality index of $\Phi_{\pi}$, which is largest integer $N$ such that there exists an $N$-dimensional subspace $M$ with the property $\Phi_{\pi}(x\otimes y) = 0$ whenever $x\perp y$ and $x, y\in M$.
 
 \item[(v)] $\max\{\beta(\Phi_{\pi}),  \lfloor {d\over 4} + 1\rfloor\}$ is a lower bound for the phase retrievability of $\Phi_{\pi}$.

 \end{itemize}

\section{Preliminaries}

We recall some notations, definitions  and basic facts that are needed for the rest of this paper.

\subsection{Notations} Here is a list of standard notations we will use in this paper:

\begin{itemize}

\item $H, K$ -- finite dimensional Hilbert spaces over $\C$, $B(H, K)$  -- the space of all the linear operators from $H$ to $K$, write $B(H) = B(H, K)$ if $H = K$. In the case that $H = \C^n$ and $K = \C^m$, $B(H, K) = M_{m\times n}(\C)$ and we use $M_n(\C)$ for the case when $m =n$. We use $I_{H}$  (or $I$ if no confusion from the context) to denote the identity operator on $H$.

\item $\langle A, B\rangle = tr(AB^*)$ is the trace inner product  on $B(H)$, and  $U(H)$ is the group of unitary operators on a complex Hilbert space $H$.

\item  For a subset $\mathcal{A}$ of $B(H)$, the commutant $\mathcal{A}' = \{T\in B(H): TA = AT, \forall A\in \mathcal{A}\}$.

\item Let $x\in H, y\in K$. We will use  $x\otimes y$ to denote the rank-one operator defined by $z \mapsto \langle z, y\rangle x$ for $z\in K$. Occasionally, $x\otimes y$  is also used  to denote the tensor product in $H\otimes K$ and the readers should be able to distinguish them from the context.

\item Let $\pi$ be a unitary representation of a group $G$, we use $m\pi$ to denote the representation $ \pi\oplus ...  \oplus \pi$ ($m$-copies). Any unitary representation $\pi$ on a finite dimensional Hilbert space can be decomposed as
$$
\pi = m_1\pi_1 \oplus ... \oplus m_d\pi_d
$$
where $\pi_i$ are inequivalent irreducible unitary representations. 

\item Two unitary representations $\pi: G\to U(H)$ and $\sigma: G\to U(K)$ are called {\it disjoint} if they have no equivalent subrepresentations., or quivalently, 
$$
Hom(\pi, \sigma) = \{T\in B(H, K): T\pi(g) = \sigma(g)T\} = \{0\}.
$$
In particular, any two inequivalent irreducible representations are disjoint.

\item $[d] = \{1, 2, ... , d\}$.

\end{itemize}

\subsection{Covariant Quantum Channels}

Let $\pi$ and $\sigma$ be unitary representations of a compact group $G$ on $\C^n$ and $\C^m$, respectively. We say that $\Phi$ is $(\pi, \sigma)$-covariant if
$$
\Phi(\pi(g)T\pi(g^{-1})) = \sigma(g)\Phi(T) \sigma(g^{-1})
$$
holds for every $g\in G$.

Covariant quantum channels form  important class of channels since many challenging problems in quantum information theory are usually more tractable when certain symmetries are imposed on the channel. We refer to for example \cite{Kirbs-03, CKKLY, DFH-Quantum, GBW, Haa1, Haa2, Haa3, MSD-JMP, Nuw, Ouyang, SDN} for some recent progresses on theoretical studies of covariant quantum channels. In particular, in their  recent work  \cite{MSD-JMP}, M.  Mozrzymas, M. Studzi'nski and N.  Datta investigated the structure of covariant quantum channels  with respect to an irreducible representation  $\pi$ for a finite group $G$, and obtained  spectral decomposition of such a covariant quantum channels  in terms of representation characteristics of the group $G$.


There is a natural way, called channel twirling,  to produce a $(\pi, \sigma)$-covariant quantum channel from any given quantum channel.  Let $\Phi: B(H) \to B(K)$ be a quantum channel, and $\pi, \sigma$ be two continuous unitary representations of a group $G$ on $H$ and $K$, respectively. Then

$$
\Psi(T) = \int_{G}\sigma(g^{-1})\Phi(\pi(g)T\pi(g^{-1}))\sigma(g) d\mu(g)
$$
is a $(\pi, \sigma)$-covariant quantum channel, where $\mu$ is the Haar measure of the compact group $G$. Note that  $$
\Psi(T) = {1\over |G|}\sum_{g\in G}\sigma(g^{-1})\Phi(\pi(g)T\pi(g^{-1}))\sigma(g)
$$
if $G$ is finite.

Now we consider a special type of covariant quantum channels (the ones twirled from the identity map): Let $\pi: G\to U(H)$ and $\sigma: G\to U(K)$ be two continuous unitary representations. We define a linear map $\Phi_{\pi, \sigma}: B(K, H)\to B(K, H)$ by
$$
\Phi_{\pi, \sigma}(T) =  \int_{G}\pi(g)T\sigma(g^{-1}) d\mu(g)
$$
and denote $\Phi_{\pi, \sigma}$ by $\Phi_{\pi}$ when $\pi = \sigma$. Then $\Phi_{\pi}$ is a $\pi$-covariant quantum channel which will be called a {\it $\pi$-induced twirling channel}.
Twirling channels have a long history in the quantum information literature and have numerous applications. For example, channels of this form have been used in the contexts of quantum error correction, quantum data hiding,  as well as in the study of quantum entanglement, and quantum coherence c.f \cite{BDSW, CGJKP, DLT, VW}

Here is a list of properties that will be need for the rest of this paper.

\begin{itemize}

\item  $\mathcal{A}_{\pi}' = range (\Phi_{\pi})$;

\item $\pi$ is irreducible if and only if $\Phi_{\pi}(T) ={1\over \dim H}  tr(T) I$ for every $T\in B(H)$;

\item $\Phi_{\pi, \sigma} = 0$ if and only if $\pi$ and $\sigma$ are strongly disjoint. In particular,  $\Phi_{\pi, \sigma} = 0$ when $\pi$ and $\sigma$ are inequivalent irreducible representations.

\end{itemize}

\subsection{Frames and phase-retrievability}

 Frame theory is closely related to operator valued measures and consequently to  quantum information theory. Phase retrieval property of a frame is probably the most relevant part  to quantum information theory. Recall that a sequence $\{f_j\}_{j\in \Bbb{J}}$ is called a {\it frame}  for a Hilbert space $H$ if there are two positive constant numbers  $A,B>0$ such that $$A||x||^2\leq\sum_{i\in I}|\langle x, f_i\rangle |^2\leq B||x||^2$$ holds for every $x\in H$. A frame is called a {\it tight frame} if $A= B$ and a {\it Parseval frame} if  $A= B =1$. A frame  $\{f_j\}_{j\in \Bbb{J}}$  is a Parseval frame if and only if $\sum_{j\in \Bbb{J}} f_j\otimes f_j = I$. Every frame  $\{f_j\}_{j\in \Bbb{J}}$ is similar to a Parseval frame in the sense that there is an invertible operator $S\in B(H)$ such that $\{Sf_j\}_{j\in\Bbb{J}}$ is a Parseval frame.   In the finite dimensional case, a finite sequence $\{x_i\}_{i=1}^{N}$ is a frame for $H$ if and only if $H = span \{x_i: 1\leq i\leq N\}$.

 A {\it phase retrieval frame} for a Hilbert space $H$ refers to a frame  $\{f_j\}_{j\in \Bbb{J}}$ in $H$ such that  the magnitudes of the frame coefficients $\langle x, f_j\rangle $ of a  signal $x\in H$ uniquely determines  the rank-one state $x\otimes x$. 
 More generally, a collection of operators $\{A_j\}_{\in\Bbb{J}}$ in $B(H)$ is called a {\it phase retrievable operator -valued frame}  for $H$ if the phaseless measurements $\langle A_jx, x \rangle$ uniquely determines $x\otimes x$. It is obvious that a (vector-valued) frame $\{f_j\}_{j\in \Bbb{J}}$ is phase retrievable if and only if $\{f_j\otimes f_j\}_{j\in \Bbb{J}}$ is a phase retrievable  operator valued frame. A natural question is to find the minimal length of a phase retrievable frame for $\R^n$ and $\C^n$. 
 
 For an $n$-dimensional Hilbert space $H$, we will use $\mathcal{I}_n$ to denote the smallest integer $N$ such that there is a phase retrievable frame $\{x_{j}\}_{j=1}^{N}$ for $H$. The following is well-known in the literature.

\begin{prop} \label{prop-2.1} If $H$ be an $n$-dimensional complex Hilbert space,  then $\mathcal{I}_n \leq 4n-4$. Moreover every generic frame $\{f_j\}_{j=1}^{N}$ of length $N\ge 4n-4$ is phase retrievable. 
\end{prop}

 A positive operator valued measure (POVM for short) or observables on a Hilbert space $H$  is a collection of positive  operators $\{F_i\}$ in $B(H)$ such that $\sum_{j\in \Bbb{J}} F_j = I_H$. A POVM $\{F_{j}\}_{j\in\Bbb{J}}$ is  {\it information complete} (c.f.  \cite{5Renes}) if $\{\langle x, F_jx\rangle\}_{j\in \Bbb{J}}$ uniquely determines  the pure state $x\otimes x$. In other words, an information complete POVM is a phase retrievable operator valued frame. 
  For a quantum channel $\Phi: B(H)\to B(K)$,  its adjoint $\Phi^*$ is unital and hence $\{\Phi^*(F_{j})\}_{j\in\Bbb{J}}$ is a POVM for $H$ whenever $\{F_{j}\}_{j\in\Bbb{J}}$ is a POVM for $K$. In the Heisenberg picture of quantum channles, a POVM in $K$ are the observables that are used to measure a state $\rho$ in $B(H)$ with measurement $\langle \rho, \Phi^*(F_j)\rangle = tr(\rho\Phi^*(F_j)) = tr(\Phi(\rho)F_j)$.
  
   It is important that a quantum channel $\Phi$ admits a POVM  on $K$  that distinguishes the pure states from $H$ (c.f. \cite{7Ariano, 6Tumalka}). 
  Such a quantum channel was called in \cite{KaiLiu-2}  {\it phase retrievable}, and some characterizations were discussed in terms of the Kraus operators.  Clearly many quantum channels are not phase retrievable. For this reason, we introduce  the following definition: A subspace $M$ of $H$  is called {\it phase retrievable} under $\Phi$  if $\Phi$ is pure state injective on $M$, i.e., $\Phi(x\otimes x) = \Phi(y\otimes y)$ implies that $x\otimes x = y\otimes y$ for $x, y\in M$, or equivalently if there exists a  POVM $\{F_{j}\}_{j\in\Bbb{J}}$ in $B(K)$ such that $\{P_{M}\Phi^*(F_{j})P_{M}\}_{j\in\Bbb{J}}$ is a phase retrievable operator valued frame for $M$, where $P_M$ is the orthogonal projection onto $M$.
The {\it  phase retrievability index}  $pr(\Phi_{\pi})$ is defined to be the largest integer $k$ such that there exits a $k$-dimensional subspace $M\subset H$ such that $M$ is phase retrievable under $\Phi$. We will examine $pr(\Phi_{\pi})$ in section 5.

\section{Quantum codes and independence numbers} 

Let $\Phi: B(H) \to B(K)$ be a quantum channel. Recall that a quantum code $\mathcal{C}$ for a noise quantum channel $\Phi$  is a subspace of the Hilbert space such that there exists another channel $\Psi$ such that
$$
\rho = \Psi \circ \Phi (\rho)
$$
for any state $\rho$ supported on $\mathcal{C}.$   In this case we say that $\mathcal{C}$ is correctable under the noise channel
$\Phi$.

\begin{lem} \label{lem-3.1} Let $\mathcal{C}$ be a subspace of $H$, and let
$P$  be the orthogonal projections  onto $\mathcal{C}$. Suppose $\Phi$  is a quantum channel  with Kraus operators $\{E_i\}_{i=1}^{r}$. Then the following are equivalent:

\begin{itemize} 
 \item[(i)] $\mathcal{C}$ is a quantum code for $\Phi$.

\item[(ii)]
 there exists a Hermitian matrix $A = [a_{ij}]$ such that 
$
PE_i^*E_jP = a_{ij}P
$
holds for all $i, j$.

\item[(iii)] For any  orthonormal basis $\{x_{k}\}_{k=1}^{m}$ of  $\mathcal{C}$,  $x_{k}\otimes x_{\ell} \perp E_{i}^*E_j$ for all $i, j$ and all $k\neq \ell$.


\item[(iv)] For any orthonormal basis $\{x_{k}\}_{k=1}^{m}$ of  $\mathcal{C}$, $\Phi(x_k\otimes x_k) \perp \Phi(x_\ell \otimes x_\ell)$ for any $k\neq \ell$.

\end{itemize}

\end{lem} 
\begin{proof} The equivalence of (i),  (ii) and (iii) are well-known (c.f. Theorem 5.2 \cite{Kribs-1}). 

(iii) $\Leftrightarrow$ (iv): Note that
$$
\langle \Phi(x_k\otimes x_k), \Phi(x_\ell\otimes _\ell) \rangle = \sum_{i, j =1}^{r} tr((E_ix_k\otimes E_ix_k)(E_jx_\ell\otimes E_jx_\ell))
$$
and $tr((E_ix_k\otimes E_ix_k)(E_jx_\ell\otimes E_jx_\ell))\geq 0$. 
Therefore 
$\langle \Phi(x_i\otimes x_i), \Phi(x_j\otimes x_j) \rangle = 0$ if and only if $$ |\langle E_jx_\ell, E_ix_k\rangle|^2 = tr((E_ix_k\otimes E_ix_k)(E_jx_\ell\otimes E_jx_\ell)=  0.$$
%
%
\end{proof} 

Related to the quantum code is the concept of {\it independence number}  for a quantum channel $\Phi$, which is the largest integer $m$ such that there is an orthonormal set $\{x_k\}_{k=1}^{m}$ such that $x_{k}\otimes x_{\ell} \perp E_{i}^*E_j$ for all $i, j$ and all $k\neq \ell$, or equivalently by Lemma \ref{lem-3.1} $\Phi(x_k\otimes x_k)$ and $ \Phi(x_\ell \otimes x_\ell)$  are orthogonal in the trace inner product for any $k\neq \ell$,  where $\{E_i\}_{i=1}^{r}$ are Kraus operators of $\Phi$. This is the same largest integer $m$ such that  Define 
with which there exist a set of states $\rho_1, ... \rho_m\in B(H)$ such that $\Phi(\rho_1) , ... , \Phi(\rho_m)$
can be perfectly distinguished c.f. \cite{DSW}.  In what follows the independence number of $\Phi$ will be denoted by $\alpha(\Phi)$. The {\it zero-error capacity} of a channel $\Phi$ is define in an asymptotic setting by
$$
\mathcal{C}_{0}(\Phi) = \lim_{n\to \infty}{1\over n} \log  \alpha(\Phi^{\otimes n}),
$$
where $\Phi^{\otimes n}$ is the $n$-fold quantum channel defined on the $B(H^{\otimes n})$. It is well known that
the zero error capacity is even harder to compute than the independence number. In fact, it is
not even known if it is in general a computable quantity in the sense of Turing c.f. \cite{Paulsen}.

\vspace{3mm}

If  $\mathcal{C}$ is a quantum code for $\Phi$, then Lemma \ref{lem-3.1} implies that $\alpha(\Phi) \geq \dim \mathcal{C}$, and hence 
$$
\alpha(\Phi) \geq \max \{  \dim \mathcal{C}:  \mathcal{C} \text{ is a quantum code for}\  \Phi\}.
$$
In what follows we will use $\beta(\Phi)$ to denote the right hand side of the above inequality. The following simple example shows that the equality does not hold in general.

\begin{exam} Let $\Phi: M_{2\times 2}(\C) \to M_{2\times 2}(\C)$ be a quantum channel with Kraus operators $E_1 = e_1\otimes e_1$ and $E_2 = e_2\otimes e_2$. Then $\Phi(e_1\otimes e_1) \perp \Phi(e_2\otimes e_2)$, and hence $\alpha(\Phi) = 2$. However, $\C^2$ is not a correctable quantum code for $\Phi$ since condition (ii) in Lemma \ref{lem-3.1} is not satisfied.  Thus $\beta(\Phi) = 1$.
\end{exam}

On the other hand there are plenty of quantum channels when the equality holds.

\begin{exam} Let $\Phi: B(H)\to B(H\oplus H)$ be a quantum channel with Kraus operators $E_1, E_2$ defined by $E_1x = {1\over \sqrt{2}} (x\oplus 0)$ and  $E_2x = {1\over \sqrt{2}} (0\oplus x)$ for any $x\in H$. Then $E_1^*E_1 = E_2E_2^*  = {1\over 2} I_{H}$, and $E_{i}^*E_{j} = 0$ if $i\neq j$. Thus $H$ is a quantum code for $\Phi$, and hence $\alpha(\Phi) = \beta(\Phi) = \dim H$.
\end{exam}

It is an interesting question to explore necessary and/or  sufficient conditions under which the equality hold. We will prove that the equality holds for twirling channel $\Phi_{\pi}$ if and only if $\pi$ is unitarily equivalent to $m\sigma$ for some irreducible representation $\sigma$ and $m\in\N$. We first show that $\alpha (\Phi_{\pi}) = \sum_{i=1}^{d}m_i$.

\begin{theo} \label{thm-IN} Suppose that $\pi=m_1\pi_1 \oplus  \cdots \oplus m_d\pi_d$  be a unitary representation of $G$ on a Hilbert space $H$, where each $\pi_i$ is irreducible and $\pi_i, \pi_j$ are inequivalent for $ \forall 1\leq i\neq  j\leq d$. Then $\alpha(\Phi_\pi)=\sum_{i=1}^d m_i$.
\end{theo}
\begin{proof} Let $\mathcal{A}_{\pi}$ be the C*-algebra generated by $\pi(G)$. Then $$\mathcal{A}_{\pi} = (I_{m_1}\otimes B(H_1))\oplus \cdots \oplus (I_{m_d}\otimes B(H_d)), $$ where $I_{m_i}$ is the identity matrix on $\C^{m_i}$. Let $\{e_{ij}\}_{j=1}^{m_i}$ be the canonical orthonormal basis for $\C^{m_{i}}$ and pick a unit vector $x_i\in H_i$. Set $x_{ij} = e_{ij}\otimes x_i$ viewing it as a vector in $H$ by considering $\C^{m_i}\otimes H_i$ as a subspace of $H$. Then it is obvious that $\langle x_{ij}, Ax_{k\ell}\rangle = 0$ for all $(i, j) \neq (k, \ell)$ and all $A\in \mathcal{A}_{\pi}$. This implies that $\alpha(\Phi_{\pi}) \geq \sum_{i=1}^dm_i$.

Conversely, suppose  $x_1, ... , x_N$ is a collection of nonzero vectors in $H$ such that  $\langle x_i, Ax_k\rangle = 0$ for all $i\neq k$ and  for all  $A\in \mathcal{A}_{\pi}$. Let $M_i = \mathcal{A}_{\pi}x_i$. Then we have that $M_i\perp M_j$ for all $i\neq j$ and  each $M_i$ is $\pi$-invariant. Let $\sigma_i$ be the restriction of $\pi$ to $M_i$. Then each $\sigma_i$ is a unitary representation and $\sigma_1\oplus ... \oplus \sigma_N$ is a subrepresentation of $\pi$. Since $\pi$ is the direct sum  of only $m_1 + ... + m_d$ number of irreducible subrepresentations, we get that $N \leq \sum_{i=1}^{d}m_i$ which implies that $\alpha(\Phi_{\pi}) \leq \sum_{i=1}^{d}m_i$. Thus we proved the claim that $\alpha(\Phi_{\pi}) = \sum_{i=1}^{d}m_i$.
%
\end{proof}

To prove $\beta(\Phi_{\pi}) = \max\{m_1, ... , m_d\}$ we first consider the following special case.

\begin{lem} \label{lem-3.2} If  $\pi = m \sigma = I_m\otimes \sigma$ acting on $\C^{m}\otimes K$ such that $\sigma: G\to  U(K)$ is irreducible, then  $\alpha(\Phi_{\pi}) = \beta(\Phi_{\pi}) = m$ 
\end{lem} 

\begin{proof}  First, by Theorem \ref{thm-IN},  we know that  $\alpha(\Phi_{\pi}) = m$. Now fix a unit vector $x\in K$ and let $x_i = e_i\otimes x$, where $\{e_i\}_{i=1}^{m}$ is the standard orthonormal basis for $\C^m$. Let $\mathcal{C} = span \{x_i\}_{i=1}^{m} = \C^{m}\otimes x$. It is enough to show that $\mathcal{C}$ is a quantum code for $\Phi$. For any $u = \vec{c}\otimes x, v = \vec{d}\otimes x\in \mathcal{C}$ such that $u\perp v$, we have that  $\vec{c}\perp \vec{d}$. Since $\mathcal{A}_{\pi} = I_{m}\otimes B(K)$, we get
$$
\langle u, Av\rangle =\langle \vec{c}, \vec{d}\rangle \cdot \langle x, Tx\rangle = 0
$$
for any $A = I_m \otimes T\in \mathcal{A}_{\pi}$, which implies by Lemma \ref{lem-3.1} that  $\mathcal{C}$ is a quantum code. Thus we obtain  $\alpha(\Phi_{\pi}) = \beta(\Phi_{\pi})$.
\end{proof}

\begin{theo}\label{thm-QC} Suppose that $\pi=m_1\pi_1 \oplus  \cdots \oplus m_d\pi_d$  be a unitary representation of $G$ on a Hilbert space $H$, where $\pi_i$ is irreducible and $\pi_i, \pi_j$  are inequivalent for $ \forall 1\leq i\neq j\leq d$. Then $\beta(\Phi_\pi) = \max \{m_1, ... , m_d\}$. 
\end{theo} 

\begin{proof}  
Let $\mathcal{C}$ be a quantum code of dimension $N$ for $\Phi_{\pi}$.  Let $\{u_j\}_{i=1}^{N}$ be an orthonormal basis for $\mathcal{C}$, and  $P_i$ be the orthogonal projection onto the subspace $\C^{m_i}\otimes H_i$. Since $P_1 + ... + P_d = I$, there exists an $i$ such that $P_iu_1 \neq 0$. For any fixed index $j\geq 2$,   $u_1 + u_j,  u_1 - u_j$ are two orthogonal vectors in $\mathcal{C}$.  Since $\mathcal{C}$ is a quantum code, we get that 
$u_1 \perp \mathcal{A}_{\pi} u_j$ and $u_1+ u_j \perp \mathcal{A}_{\pi}(u_1-u_j)$. In particular since $P_i\in \mathcal{A}_{\pi}$ we have
$$
\langle u_j, P_iu_1 \rangle  =0 \ \ \text{and} \ \ \langle u_1 + u_j , P_i(u_1 - u_j) \rangle  =0.
$$
The above two combined to  imply that $Pu_j\perp Pu_1$ and $||P_iu_j|| = ||P_iu_1||$. With the same argument by replacing $u_1$ by $u_j$, and $j$ by another index $j'$, we clearly get that $\{P_iu_j\}_{j=1}^{N}$ is an orthogonal set of nonzero vectors in $\C^{m_i}\otimes H_i$ such that $P_iu_j \perp \mathcal{A}_{m_i\pi_i}P_iu_{j'}$ for any $j\neq j'$.  Thus $N\leq \alpha(\Phi_{m_i\pi})=  m_i$, and therefore $\beta(\Phi_{\pi}) \leq \max\{m_i : 1\leq i\leq d\}$.

On the other hand, without losing the generality we can assume that $m_1 = \max \{m_i: i=1,... , d\}$. By Lemma \ref{lem-3.2}, there is a  $m_1$-dimensional quantum code $\mathcal{C}_{1}$  in $\C^{m_1}\otimes H_1$ for $\Phi_{m_1\pi_1}$. Clearly $\mathcal{C} = \mathcal{C}_{1} \oplus 0\oplus ... \oplus 0$ is a quantum code of $\Phi_{\pi}$. Thus we have $\beta(\Phi_{\pi}) \geq \dim \mathcal{C} = m_1$, and consequently we have proved  $\beta(\Phi_\pi) = \max \{m_1, ... , m_d\}$.
\end{proof}

%
%

\begin{coro} \label{coro-3.1}  Let $\pi=m_1\pi_1 \oplus  \cdots \oplus m_d\pi_d$  be a unitary representation of $G$ on a Hilbert space $H$, where $\pi_i$ is irreducible and $\pi_i, \pi_j$ are inequivalent for $ \forall 1\leq i\neq jj\leq d$. Then   $\alpha(\Phi_{\pi}) = \beta(\Phi_{\pi})$ if and only if $d=1$.
\end{coro} 
\begin{proof} If $d=1$, then $\alpha(\Phi_{\pi}) = \beta(\Phi_{\pi})$ follows from Lemma \ref{lem-3.2}. Conversely, since $\alpha(\Phi_{\pi}) = m_1+ ... + m_d$ and $ \beta(\Phi_{\pi}) = \max\{m_1, ... , m_d\}$, we immediately get $d =1$ if $\alpha(\Phi_{\pi}) = \beta(\Phi_{\pi})$.
\end{proof}

Let $G$ be a group and $\pi=m_1\pi_1 \oplus  \cdots \oplus m_d\pi_d$  be a unitary representation of $G$ onto a finite dimensional Hilbert space $H$. Let $G^{n} = \{\vec{g} =(g_1, ... , g_n): g_i\in G\}$ be the product group and $\pi_{\pi}^{\otimes n}$ be the unitary presentation of $G$ on $H^{\otimes n}$ defined by
$$
\pi_{\pi}^{\otimes n}(\vec{g}) =  \pi(g_1)\otimes \cdots \otimes \pi(g_n), \ \ \forall \vec{g}\in G^{n}.
$$

\begin{theo} \label{thm-capasity} Let $\pi=m_1\pi_1 \oplus  \cdots \oplus m_d\pi_d$  be a unitary representation of $G$ on a Hilbert space $H$, where $\pi_i$ is irreducible and $\pi_i, \pi_j$ are inequivalent for $ \forall 1\leq i\neq j\leq d$. Then  
$$
 \mathcal{C}_{0}(\Phi_{\pi}) = \log (\sum_{i=1}^{d} m_i).
$$
\end{theo} 
\begin{proof} Write $[d]^{n} = \{ (k_1, ... , k_n): k_i\in [d]\}$. The $\pi_{\pi}^{\otimes n}$ has the decomposition of  the form:
$$
\pi_{\pi}^{\otimes n} = \sum_{(k_1, ..., k_n)\in [d] ^{n}}^{\oplus} m_{k_1}\cdots m_{ k_n} (\pi_{k_1}\otimes \cdots \otimes \pi_{k_n})
$$
Note that  $\pi_{k_1}\otimes \cdots \otimes \pi_{k_n}$ is irreducible,  and  $\pi_{k_1}\otimes \cdots \otimes \pi_{k_n}$ and  $\pi_{k'_1}\otimes \cdots \otimes \pi_{k'_{n}}$ are inequivalent whenere $(k_1, ... , k_n) \neq (k'_1, ... , k'_n)$ (This can be easily checked by comparing their characters). Thus,  by Theorem \ref{thm-IN}, we get 
$$
\alpha(\Phi_{\pi}^{\otimes n}) = \alpha(\Phi_{\pi^{\otimes n}}) = \sum_{(k_1, ..., k_n)\in [d]^{n}} m_{k_1} m_{ k_2}\cdots m_{k_n} = (m_1+ ... +m_d)^n
$$
and hence  $\mathcal{C}_{0}(\Phi_{\pi}) = \lim_{n\to\infty}{1\over n} \log \alpha(\Phi_{\pi}^{\otimes n}) = \log (\sum_{i=1}^{d} m_i)$.
\end{proof}

\section{Orthogonality index of $\Phi_{\pi}$}

While the independent number is the largest integer $\alpha(\Phi)$ such that there exists an orthonormal set $\{x_i\}_{i=1}^{N}$ with the property 
$\Phi(x_i\otimes x_i) \perp \Phi(x_j\otimes x_j)$ for any $i\neq j$, we define the {\it orthogonality index} of $\Phi$, denote it by $\gamma(\Phi)$,  to be the largest number $N$ such that there exists  $\{x_i\}_{i=1}^{N}$ with the property  $\Phi(x_i\otimes x_j) = 0$ for any $i\neq j$. This is a concept  related to strongly disjoint frames that plays extremely important roles in frame theory and in establishing a Balian-Low type of duality principle for group representation frames c.f. \cite{BDLL-JFAA, DHL-JFA, Han-Larson-AMSM, Han-Larson-BLMS}.

Let $\{x_i\}_{i=1}^{N}$ be a sequence in a Hilbert space $H$ and and $\{y_i\}_{i=1}^{N}$ be a sequence in a Hilbert space $K$. We say that
$\{x_i\}_{i=1}^N$ $\{y_i\}_{i=1}^N$ for  are {\it strongly disjoint} if 
$
\sum_{i=1}^{N}\langle x, x_i\rangle y_i = 0
$
for all $x\in H$, or equivalently,  $\sum_{i=1}^Ny_i\otimes x_i = 0$. Consequently, $\Phi_{\pi}(x\otimes y) = 0$ if and only if $\{\pi(g)x\}_{g\in G}$ and $\{\pi(g)x\}_{g\in G}$ are strongly disjoint. In this case we also say that $x$ and $y$ are {\it $\pi$-orthogonal}  \cite{DHL-JFA}.

\begin{lem}\label{lem-4.1} Let $\pi: G \to U(H)$ be a unitary representation and $x, y\in H$. Then $\Phi_{\pi}(x\otimes y) = 0$ if and only if $\mathcal{A}_{\pi}'x \perp \mathcal{A}_{\pi}'y$.
\end{lem}
\begin{proof}  Note $\mathcal{A}_{\pi}' = \{\Phi(T): T\in B(H)\}$ and $\Phi = \Phi^*$. Thus we have
$$
\langle T,  \Phi(x\otimes y) \rangle = \langle \Phi(T), x\otimes y \rangle = \langle \Phi(T)x, y\rangle
$$
which implies that  $ \Phi(x\otimes y) = 0$ if and only if $x\perp \Phi(T)y$ for every $T\in B(H)$. Thus we get that $x$ and $y$ are $\pi$-orthogonal if and only if $\mathcal{A}_{\pi}'x \perp \mathcal{A}_{\pi}'y$.
\end{proof}

\begin{lem} \label{lem-4.2} If  $\pi = m \sigma: G \to U(H)$ acting on $H = \C^{m}\otimes K$ such that $\sigma: G\to  U(K)$ is irreducible, then $\gamma(\Phi_{\pi}) = \dim K$.
\end{lem}
\begin{proof} If $\Phi_{\pi}(x\otimes y) =0$ for some $x, y\in H = \C^{m}\otimes K$, then by Lemma 4.1 we have $\mathcal{A}_{\pi}'x \perp \mathcal{A}_{\pi}'y$, where $\mathcal{A}_{\pi}' = M_{m}(\C) \otimes I$. Note that we can always write $x, y$ in the form of $x = \sum_{i=1}^{m}e_i\otimes x_i$ and $y = \sum_{i=1}^{m} e_i\otimes y_i$ for some $x_i, y_i\in K$, where $\{e_i\}_{i=1}^{m}$ is the canonical orthonormal basis for $\C^m$. Let $E_{ii} = e_i\otimes e_i\in M_{m}(\C)$. Since $E_{ii} \otimes I \in \mathcal{A}_{\pi}'$, we get 
$$
\langle x_i, y_j \rangle =  \langle (E_{ii}\otimes I)x, (E_{jj}\otimes I) y\rangle  = 0
$$
for all $i, j \in [m]$.

Now let $\{u_i\}_{j=1}^{N}$ be an orthonormal set in $H$  such that  $\Phi_{\pi}(u_i\otimes u_j) = 0$ for any $i\neq j$. Write $u_i = \sum_{j=1}^{m} e_i\otimes u_{ij}$, where $u_{ij}\in K$. For each $i$, pick an index $n_i$ such that $u_{i n_i} \neq 0$. Then by the above argument we that $\{u_{in_i}\}_{i=1}^{N}$ is an orthogonal set of nonzero vectors in $K$. This implies that $N \leq \dim K$, and hence $\gamma(\Phi_{\pi}) \leq \dim K$. 

On the other hand, let $\{u_i\}_{i=1}^{n}$ be an orthonormal basis for $K$ and let 
 $x_i = e_i\otimes u_i \in H$ for $i\in [n]$. Then clearly we have $\mathcal{A}_{\pi}'x_i \perp \mathcal{A}_{\pi}'x_j$ for any $i\neq j$. Thus $\gamma(\Phi_{\pi}) \geq \dim K$, which completes the proof.
\end{proof} 

Now we prove for the general case.

\begin{theo} \label{thm-orth} Suppose that $\pi=m_1\pi_1 \oplus  \cdots \oplus m_d\pi_d$  be a unitary representation of $G$ on a Hilbert space $H$, where each $\pi_i$ is a irreducible  representation on $H_{i}$ and $\pi_i, \pi_j$ are inequivalent for $ \forall 1\leq i\neq j\leq d$. Then  $\gamma(\Phi_\pi)=\sum_{i=1}^k n_i$, where $n_i = \dim H_{i}$.
\end{theo}
\begin{proof} Since $\mathcal{A}_{\pi} = I_{m_1}\otimes B(H_1) \oplus \cdots \oplus I_{m_d}\otimes B(H_d)$, we get that $$\mathcal{A}_{\pi} '= M_{m_1}(\C) \otimes I_{H_1} \oplus \cdots \oplus M_{m_d}(\C)\otimes I_{H_d}.$$ By Lemma 4.1, there is an orthonormal basis $\{x_{ij}\}_{j=1}^{n_i}$ for $\C^{m_{i}}\otimes H_i$ that $$x_{ij} \perp (M_{m_i}(\C)\otimes I_{H_i})x_{ik}$$ for $1\leq i \neq k\leq m_i$. This implies that 
$$
x_{ij} \perp \mathcal{A}_{\pi}' x_{k\ell}
$$  whenever $(i, j) \neq	(k, \ell)$. Thus $\Phi_{\pi}(x_{ij}\otimes x_{k\ell}) = 0$ for all $(i, j)\neq (k, \ell)$, which implies that $\gamma(\Phi_{\pi}) \geq \sum_{i=1}^{d}n_i$.

For the other direction of the inequality,  let $N = \gamma(\Phi_{\pi})$. Then there exists an orthonormal set $\{x_j\}_{j=1}^{N}$ for $H$ such that $x_j \perp \mathcal{A}_{\pi}'x_k$ for all $j\neq k$. Let  $P_i$ be the orthogonal projection onto the subspace $\C^{m_i}\otimes H_i$. Then $P_i\in \mathcal{A}_{\pi}'$.  This implies that $P_ix_j\perp \mathcal{A}_{m_i\pi_i}'P_ix_k $ for all $k\neq \ell$ and every $i$. In particular, we have  $P_ix_j \perp P_ix_k$ for all $j\neq k$. 
Define subsets $\Lambda_1, ..., \Lambda_d$ of $\{1, ... , N\}$ inductively by 
$$
\Lambda_1 = \{j\in [N]: P_{1}x_j \neq 0\}
$$
and 
$$
\Lambda_{i} =  \{i\notin \Lambda_{i-1}: P_{i}x_j\neq 0\} 
$$
for $2\leq	 i\leq d$. Then $[N] = \cup_{i=1}^{d}\Lambda_i$. Let $y_j= P_1x_j$ for $j\in \Lambda_1$. Then $\{y_j\}_{j\in \Lambda_1}$ is a collection of nonzero orthogonal vectors such that $y_j\perp \mathcal{A}_{m_i\pi_i}'y_k $ for all $k\neq \ell$  in $\Lambda_1$. Thus, by Lemma \ref{lem-4.2}, we have $|\Lambda_1 | \leq n_1$. With the same arguments we also have  $ |\Lambda_i | \leq n_i$ for $i=2, ... , d$. Therefore we get $\gamma(\Phi_{\pi}) = N = \sum_{i=1}^{d}|\Lambda_i | \leq \sum_{i=1}^{d} n_i$, which completes the proof.
\end{proof}

Note that $\gamma(\Phi)$ can be considered as a ``dual object" of $\alpha(\Phi)$. Similarly it is natural to consider a ``dual version" of $\beta(\Phi)$.  For this let us define $\tau(\Phi)$ to be the largest integer $L$ such that there exists a $L$-dimensional subspace $M$ with the property that $\Phi(x\otimes y) = 0$ whenever $x\perp y$ and $x, y\in M$. We have the following dual theorem of Theorem \ref{thm-QC}

\begin{theo} \label{thm-4.2} Suppose that $\pi=m_1\pi_1 \oplus  \cdots \oplus m_d\pi_d$  be a unitary representation of $G$ on a Hilbert space $H = \C^{m_1}\otimes H_1 \oplus \cdots \oplus \C^{m_d}\otimes H_d$, where $\pi_i, \pi_j$  are inequivalent  irreducible representations for $ \forall 1\leq i\neq j\leq d$. Then $\tau(\Phi) = \max\{n_1, ... , n_d \}$, where $n_i = \dim H_i$ for each $i\in [d]$.
\end{theo} 

\begin{proof}  We first show that $L \geq n_i$ where $n_i = \dim H_i$. It suffices to check that $L \geq n_1$. Let $M = e_1\otimes H_1$. Then two vectors $x = e_1\otimes u, y = e_1\otimes v\in M$ are orthogonal if and only if $u$ and $v$ are orthogonal vectors in $H_1$ Thus for any $B= B_1 \otimes I_{H_1} \oplus \cdots \oplus B_d\otimes I_{H_d}\in  \mathcal{A }_{\pi}' $ we get
$
\langle x, By\rangle = \langle e_1, B_1e_1\rangle \cdot \langle u, v\rangle = 0,
$
which implies by Lemma 4.1 that $\Phi_{\pi}(x\otimes y) = 0$. Thus $L \geq n_1$, and therefore we get $L \geq \max\{n_1, ... , n_d \}$ 

The idea for proof of the other direction of the inequality is almost identical to the proof of Theorem 3.3 and we still include it here for self completeness.  Let $M$ be a subspace such that $\Phi_{\pi}(x\otimes y) = 0$ whenever $x, y\in M$ are orthogonal vectors.  Let $\{u_j\}_{i=1}^{N}$ be an orthonormal basis for $M$, and  $P_i$ be the orthogonal projection onto the subspace $\C^{m_i}\otimes H_i$. Since $u_1 \neq 0$ and  $P_1 + ... + P_d = I$, there exists an index $i$ such that $P_iu_1 \neq 0$. For any fixed index $j\geq 2$,   $u_1 + u_j,  u_1 - u_j$ are two orthogonal vectors in $M$. Thus 
$u_1 \perp \mathcal{A}_{\pi}' u_j$ and $u_1+ u_j \perp \mathcal{A}_{\pi}'(u_1-u_j)$. Since $P_i\in \mathcal{A}_{\pi}'$ we get
$$
\langle u_j, P_iu_1 \rangle  =0 \ \ \text{and} \ \ \langle u_1 + u_j , P_i(u_1 - u_j) \rangle  =0.
$$
The above two combined to  imply that $Pu_i\perp Pu_1$ and $||P_iu_j|| = ||P_iu_1||$. With the same argument by replacing $u_1$ by $u_j$, and $j$ by another index $j'$, we clearly get that $\{P_iu_j\}_{j=1}^{N}$ is an orthogonal set of nonzero vectors in $\C^{m_i}\otimes H_i$ such that $P_iu_j \perp \mathcal{A}_{m_i\pi_i}'P_iu_{j'}$ for any $j\neq j'$.  Thus $N\leq \gamma(\Phi_{m_i\pi})$, and hence it follows fromTheorem 4.3 that  $N \leq m_i$. Therefore we get $N \leq \max\{n_i : 1\leq i\leq d\}$, which completes the proof.
\end{proof}

\section{Phase-retrievability of $\Phi_{\pi}$}

 \begin{lem}\label{lem-5.1} \cite{KaiLiu-2} Let  $\Phi: B(H)\to B(K)$ be a quantum channel and $M$ be a subspace of $H$. The the following are equivalent:
 \begin{itemize}
 \item[(i)] There exists a POVM $\{F_j\}_{j=1}^{N}$ such that $\{ \langle x,\Phi^{*}(F_j)x\rangle\}_{j=1}^{N}$ uniquely determines $x\otimes x$ for every $x\in M$ ( In this case we say that $\Phi$ is phase retrievable on $M$).
 
 \item[(ii)] $x\otimes x = y\otimes y$ whenever $\Phi(x\otimes x) = \Phi(y\otimes y)$ and $x, y\in M$ (In this case we say that $\Phi$ is pure state injective on $M$).
 
 \end{itemize}
 \end{lem} 
 
 Recall that   $pr(\Phi_{\pi})$ the the largest integer $k$ such that there exits a $k$-dimensional subspace $M\subset H$ such that $M$ is phase retrievable under $\Phi$.

\begin{lem} \label{lem-5.2} Suppose that $\pi=m_1\pi_1 \oplus  \cdots \oplus m_d\pi_d$  be a unitary representation of $G$ on a Hilbert space $H$, where each $\pi_i$ is an irreducible representation on $H_i$,  and $\pi_i, \pi_j$  are inequivalent for $ \forall 1\leq i\neq j\leq d$. Then  $pr(\Phi_{\pi}) \geq \max\{ m_1, ... , m_d\} = \beta(\Phi_{\pi})$. 
\end{lem}

\begin{proof}  It suffices to show that $pr(\Phi_{\pi}) \geq m_1$.  Fix a unit vector $u\in H_1$ and let $M = \C^{m_1}\otimes x$. Then for any $x = \vec{a}\otimes u$, we can write the rank-one operator $x\otimes x$ in the matrix form
$$
x\otimes x = [a_ix\otimes a_jx]_{m_1\times m_1},
$$
where $\vec{a} = (a_1, ... , a_{m_1})$.
Thus $\Phi_{\pi}(x\otimes x) =  [\Phi_{\pi_1}(a_iu\otimes a_ju)]_{m_1\times m_1}.$ Since $\pi_1$ is irreducible, we know that $\Phi_{\pi_1}(T) = {1\over |G|} tr(T) I_{H_1}$. Therefore we get that 
$$
\Phi_{\pi}(x\otimes x) =  [\Phi_{\pi_1}(a_iu\otimes a_ju)]_{m_1\times m_1} =  [a_i\bar{a}_{j} I_{H_1}]_{m_1\times m_1}.
$$

Now if  $\Phi_{\pi}(x\otimes x) = \Phi_{\pi}(y\otimes y)$ for two vectors $x = \vec{a}\otimes u, y = \vec{b}\otimes u\in M$, then we have  $ a_i\bar{a}_{j} =  b_i\bar{b}_{j}$ for all $i, j = 1, ... , m$, which implies that $x\otimes x = y\otimes y$. Therefore $pr(\Phi_{\pi}) \geq m_1$. 
\end{proof}


\begin{exam}  $pr(\Phi_{\pi}) = 1$ if either (i)  $\pi$ is irreducible or  (ii) $\pi = \pi_{1} \oplus \pi_2$ such that $\pi_1$ and $\pi_1$ are inequivalent one-dimensional representation.
\end{exam}
\begin{proof} (i) follows from the fact that $\Phi_{\pi}(x\otimes x) = \Phi_{\pi}(x\otimes x)$ whenever $||x|| = ||y||$. Thus any subspace of dimension greater than 1 can not be phase retrievable for $\Phi$.

For (ii), it is sufficient to point out that $\Phi_{\pi}$ is not pure state injective on $H = H_1\oplus H_2$. Pick two unit vectors $x_1\in H_1, x_2\in H_2$. Let $x = x_1\oplus x_2$ and $y = x_1 \oplus (-x_2)$. Then 
$\Phi_{\pi}(x\otimes x) = I_{H_1}\oplus I_{H_2} = \Phi_{\pi}(y\otimes y)$. However, $x\otimes x \neq y\otimes y$. Thus $\Phi_{\pi}$ is not pure state injective on $H$.
\end{proof}

%
%
%

Our goal is to get an reasonable estimate for $pr(\Phi_{\pi})$. For this purpose we  present  the following characterization of phase retrievable subspaces for representations of multiplicity one i.e., $m_1 = .... = m_d =1$.

\begin{prop} \label{prop-5.1} Let $\pi = \pi_1\oplus ... \oplus \pi_d$ be a unitary representation of $G$  acting on $H = H_1\oplus \cdots \oplus H_d$ such that $\pi_i$ and $\pi_j$ are inequivalent irreducible representations for all $i\neq j$, and let $M$ be a $k$-dimensional subspace of $H$. . Then the following are equivalent:
\begin{itemize}

\item[(i)] $M$ is phase retrievable for $\Phi_{\pi}$

\item[(ii)]  $M = range (T)$ for some linear map $T = (T_1, ... , T_d): \C^k \to H$ such that $\{T_i^*T_i\}_{i=1}^{d}$ does phase retrieval for $\C^k$, where $T\xi = T_1\xi \oplus ... \oplus T_d\xi$ for all $\xi\in\C^k$.
\end{itemize}

\end{prop}

\begin{proof} (i) $\Rightarrow$ (ii): Let $P_i$ be the orthogonal projection from $H$ to $H_i$ for $i=1, ... , d$, and $U: \C^k \to M$ be a unitary map. Define $T_i = P_iU$ and $T = (T_1, ... , T_d)$. Then 
$$
T\xi = T_1U\xi \oplus \cdots \oplus T_d\xi = P_1U\xi \oplus 
\cdots \oplus P_d U\xi  = (P_1\oplus \cdots \oplus P_d)U\xi = U\xi.
$$
Thus $ range(T) = M$. Suppos that $\xi, \eta\in \C^k$ such that $$\langle \xi \otimes \xi, T_i^*T_i\rangle = \langle \eta \otimes \eta , T_i^*T_i\rangle, \  \  i.e., \ \  ||T_i\xi||^2 = ||T_i\eta||^2$$ for every $i\in [d]$.  Since $\pi_1, ... , 
\pi_d$ are inequivalent irreducible representations we get that 
$$
\Phi_{\pi_i, \pi_j}(T_{i}\xi\otimes T_j\xi) = 0
$$
if $i\neq j$ and $\Phi_{\pi_i}(T_{i}\xi\otimes T_iU\xi) ={1\over \dim H_i} ||T_i\xi||^2 I_{H_i}$. Thus we get
$$
\Phi_{\pi}(T\xi\otimes T\xi) = [\Phi_{\pi_i, \pi_j}(T_{i}\xi\otimes T_j\xi) ]_{d\times d} = [\Phi_{\pi_i, \pi_j}(T_{i}\eta\otimes T_j\eta)]_{d\times d} = \Phi_{\pi}(T\eta\otimes T\eta).
$$
Since $T\xi, T\eta\in M$ and  $\Phi_{\pi}$ is pure state injective on $M$, we get that $T\xi\otimes T\xi= T\eta\otimes T\eta$, which implies that $\xi\otimes \xi= \eta\otimes \eta$. Therefore $\{T_i^*T_i\}_{i=1}^{d}$ does phase retrieval for $\C^k$.

(ii) $ \Rightarrow $ (i):  Let $x = T\xi, y = T\eta\in M$ be such that $\Phi_{\pi}(x\otimes x) = \Phi_{\pi}(y\otimes y)$. Then we get $\Phi_{\pi_i, \pi_j}(T_{i}\xi\otimes T_j\xi) = \Phi_{\pi_i, \pi_j}(T_{i}\eta\otimes T_j\eta)$ for all $i, j\in [d]$, which implies that 
$$
||T_i\xi||^2 = ||T_j\eta||^2, i.e., \langle  x, T_i^*T_ix \rangle = , \langle y, T_i^*T_iy \rangle
$$ 
for every $i\in [d]$. Since  $\{T_i^*T_i\}_{i=1}^{d}$ does phase retrieval for $\C^k$, we get that $\xi\otimes \xi = \eta\otimes \eta$ which in turn implies that   $x\otimes x = y\otimes y$. Therefore 
$M$ is phase retrievable for $\Phi_{\pi}$.
\end{proof}

\begin{coro} \label{coro-5.1} Let $\pi = \pi_1\oplus ... \oplus \pi_d$ be a unitary representation of $G$  acting on $H = H_1\oplus \cdots \oplus H_d$ such that $\pi_i$ and $\pi_j$ are inequivalent irreducible representtaions for all $i\neq j$. Then $pr(\Phi_\pi) \geq  \lfloor {d \over 4} + 1\rfloor$.
\end{coro}
\begin{proof} Let $k =  \lfloor {d \over 4} + 1\rfloor$. Then $d\geq 4k-4$, and hence by Proposition \ref{prop-2.1} there exists a phase retrievable frame $\{\xi_i\}_{i=1}^{d}$ for $\C^k$. Since $\{S\xi_i\}_{i=1}^{d}$ is also a phase  retrievable frame for any invertible matrix $S\in M_{k\times k}(\C)$, we can assume that $\xi_i = e_i$ for $i\in [k]$, where $\{e_1, ... , e_k\}$ is the canonical orthonormal basis for $\C^k$. Pick unit vectors $x_i\in H_i$ for $i\in [d]$, and let $T_i = x_i\otimes \xi_i: \C^k \to H_i$ be the rank-one operator defined by $T_i\xi =  \langle \xi, \xi_i\rangle x_i \ (\forall \xi\in\C^k)$. We claim that $T = (T_1, ... , T_d): \C^k \to H$ is a rank-k linear operator. Clearly it is enough to show that $Te_1, ... , Te_k$ are linearly independent. Suppose that $ \sum_{i=1}^{k}c_iTe_i =0$ for some scalars $c_i\in \C$. Note that
 \begin{align*}
 \sum_{i=1}^{k}c_iTe_i &= \sum_{i=1}^{k}c_i\langle e_i, \xi_1\rangle x_1 \oplus \cdots \oplus \sum_{i=1}^{k}c_i\langle e_i, \xi_d\rangle x_d \\
 &= c_1x_1\oplus \cdots \oplus c_kx_k \oplus \sum_{i=1}^{k}c_i\langle e_i, \xi_{k+1}\rangle x_{k+1} \cdots \oplus  \sum_{i=1}^{k}c_i\langle e_i, \xi_d\rangle x_d.
 \end{align*}
 Thus we get $c_1 = ... = c_k=0$, and hence $Te_1, ... , Te_k$ are linearly independent. Now let $M = range (T)$. Then $M$ is a $k$-dimensional subspace of $H$.  Since $T_i^*T_i = \xi_i\otimes \xi_i$ and $\{\xi_i\}_{i=1}^{d}$ is a phase retrievable frames for $\C^k$, we immediately get from Proposition \ref{prop-5.1} that  $M$ is phase retrievable for $\Phi_{\pi}$, and therefore  $pr(\Phi_\pi) \geq k \geq  \lfloor {d \over 4} + 1\rfloor$.
\end{proof}

Since it is easy to see by definition that $pr(\Phi_{\pi}) \geq pr(\Phi_{\sigma})$ if $\sigma$ is a subrepresentation of $\pi$, we get the following  lower bound of $pr(\Phi_{\pi})$ from Lemma \ref{lem-5.2} and Corollary \ref{coro-5.1}.

\begin{theo} Let $\pi = m_1\pi_1\oplus ... \oplus m_d\pi_d$ be a unitary representation of $G$  such that $\pi_i$ and $\pi_j$ are inequivalent irreducible representations for all $i\neq j$. Then $$pr(\Phi_\pi) \geq  \max \{\beta(\Phi_{\pi}),  \lfloor {d \over 4} + 1\rfloor\}.$$
\end{theo}


Recall that $\mathcal{I}_{k}$ is the smallest integer such that there is a phase retrievable frame of $\mathcal{I}_{k}$ vectors for $\C^k$. Now consider the case when  $\pi = \pi_1\oplus ... \oplus \pi_d$ such that  $\pi_i$ and $\pi_j$ are inequivalent one-dimensional  irreducible representations for all $i\neq j$. We claim that if $\mathcal{I}_k \leq d < \mathcal{I}_{k+1}$, then $pr(\Phi_\pi) = k =  \max \{\beta(\Phi_{\pi}),  k\}$ (Note $\beta(\Phi_{\pi}) =1$ in this case).

 Indeed, by replacing $d$ with $\mathcal{I}_{k}$ in the proof of Corollary \ref{coro-5.1}, we get $pr(\Phi_{\pi}) \geq k$. Therefore we only need to show that $pr(\Phi_\pi) \leq k$. Let $M$ be a $L$-dimensional subspace of $H$ such that $\Phi_{\pi}$ is pure state injective on $M$. Then by Proposition \ref{prop-5.1}, there exists a linear operator $T = (T_1, ... , T_d): \C^{L}\to H$ such than $range(T) = M$ and $\{T_i^*T_i\}_{i=1}^{d}$ does phase retrieval for $\C^{L}$. Since  $H_i$ is one dimensional, we know that $T_i$ is rank-one operator. Write $T_i = x_i\otimes \xi_i$ for some $x_i\in H_i$ and $\xi_i\in\C^{L}$. Then $T_{i}^*T_i = ||x_i||^{2}\xi_i\otimes \xi_i$. This implies that $\{\xi_i\}_{i=1}^{d}$ is a phase retrievable frame for $\C^{L}$, and thus $d\geq \mathcal{I}_{L}$. If $L \geq k+1$, then $\mathcal{I}_{L}\geq \mathcal{I}_{k+1}$ and this would have implied that $d\geq \mathcal{I}_{k+1}$ which is a contradiction. Thus $L \leq k$ which implies that  $pr(\Phi_\pi) \leq k$. Therefore $pr(\Phi_\pi) = k =  \max \{\beta(\Phi_{\pi}),  k\}$ for this case. We make the following conjecture:

\vspace{3mm}

\noindent{\bf Conjecture.}  Let $\pi = m_1\pi_1\oplus ... \oplus m_d\pi_d$ be a unitary representation of $G$  such that $\pi_i$ and $\pi_j$ are inequivalent irreducible representations for all $i\neq j$ and   $\mathcal{I}_k \leq d < \mathcal{I}_{k+1}$. Then $pr(\Phi_\pi) =  \max \{\beta(\Phi_{\pi}),  k\}$.


%


\begin{thebibliography}{999}
 
 
 \bibitem{BDLL-JFAA}  E. Balan, D. Dutkay, D. Han, D. Larson and F. Luef, A duality principle for groups II: Multi-frames meet super-frames, {\it J. Fourier Anal. Appl.,}   26 (6) (2020), 1-19.
 
 \bibitem{BDSW} C. Bennett, D. DiVincenzo, J.  Smolin and W.  Wootters, 
Mixed-state entanglement and quantum error correction, {\it Phys. Rev. A, } 54 (5) (1996), 3824-3851.
 
 \bibitem{Kribs-0}  S.L. Braunstein, D.W. Kribs, M.K. Patra, Zero-error subspaces for quantum channels, {\it IEEE International Symposium on Information Theory (ISIT)}, (2011), 104-108.
 
 \bibitem{CGJKP}  J. Chen, S.  Grogan, N. Johnston, C.  Kwong and  Plosker, 
 Quantifying the coherence of pure quantum states, 
{\it Phys. Rev. A,}  94 (4) (2016).
 
 \bibitem{Kirbs-03} D.W. Kribs, Quantum channels, wavelets, dilations, and representations of $O_n$, {\it Proceedings of the Edinburgh Mathematical Society,} 46 (2003), 421-433
 
\bibitem{Kribs-1} D. Kribs, A quantum computing primer for operator theorists, {\it Lin. Alg. Appl.,} 400 (2005), 147 - 167.

\bibitem{Kribs-2} D.W. Kribs, C. Mintah, M. Nathanson, R. Pereira, Quantum error correction and one-way LOCC, {\it Journal of Mathematical Physics, }60 (2019), 032202.



\bibitem{Twirling} M.  Girard and J.  Levick, Twirling channels have minimal mixed-unitary rank,  {\it Lin. Alg. Appl.,}  615( 2021), 207-227.



\bibitem{CKKLY}  E. Chang, J.  Kim, H.  Kwak, H. Lee and S.  Youn, Irreducibly  $SU(2)$-covariant quantum channels of low rank, {\it Reviews in Mathematical Physics,} to appear (arXiv: 2105.00709)


\bibitem{Choi-LAA1} M. D. Choi. Completely positive maps on complex matrices, {\it Lin. Alg. Appl.}, 10(1975), 285–290.

\bibitem{DLT} D.  DiVincenzo, D. Leung, and B. Terhal, 
Quantum data hiding,
{\it IEEE Trans. Inf. Theory,} 48 (3) (2002), 580-598.

\bibitem{Choi-LAA2} M. D. Choi, Positive semidefinite bi-quadratic forms, {\it Lin. Alg. Appl.,} 12(1975),  95-100.

\bibitem{7Ariano} G.M. d’Ariano,  P. Perinotti, and M. F. Sacchi, Informationally complete measurements and group representation, {\it Journal of Optics B: Quantum and Semiclassical Optics,}  6.6 (2004): S487.

\bibitem{DFH-Quantum} N. Datta, M. Fukuda and A.S. Holevo, Complementarity and additivity for covariant channels,  {\t Quantum Inf. Process,}  5(2), 179-207 (2006)

 \bibitem{DSW} R.  Duan, S. Severini, and A. Winter,  Zero-error communication via quantum channels,
noncommutative graphs, and a quantum Lovász number, {\it IEEE Trans. Info. Theory,}  59(2013), 1164 - 1174.



 \bibitem{DHL-JFA} D. Dutkay, D. Han and D. Larson, A duality principle for groups,  {\it J. Funct. Anal.,}   257( 2009), 1133--1143.
 



\bibitem{Paulsen}   V. Gupta, P.  Mandayam and  V.S. Sunder,  The functional analysis of quantum information theory, A Collection of Notes Based on Lectures by Gilles Pisier, K. R. Parthasarathy, Vern Paulsen and Andreas Winter, Lecture Notes in Physics 902, Springer, 2015

\bibitem{GBW}  M. Gschwendtner, A. Bluhm and A. Winter, Programmability of covariant quantum channels, {\it Quantum}, 5(2021), Page 488.

\bibitem{Haa1}  E.  Haapasalo, The Choi–Jamiołkowski isomorphism and covariant quantum channels, {\it Quantum Studies: Mathematics and Foundations,}  8(2021), 351-373.

\bibitem{Haa2} E.  Haapasalo, Compatibility of Covariant Quantum Channels with Emphasis on Weyl Symmetry, {\it Annales Henri Poincaré,}  20 (2019), 3163–3195.

\bibitem{Haa3}  E Haapasalo, JP Pellonpää,  Optimal covariant quantum measurements, {\it Journal of Physics A: Mathematical and Theoretical},  54 (2021), 155304.



\bibitem{Han-2} D. Han, T. Juste, Y. Li and W. Sun, Frame phase-retrievability and exact phase-retrievable frames, {\it  J. Fourier Anal. Appl., } 25(2019), 3154-3173.

\bibitem{Han-3} D. Han and T. Juste, Phase-retrievable operator-valued frames and representations of quantum channels, {\it  Lin. Alg. Appl.,} 579(2019), 148-168.

\bibitem{Han-Larson-AMSM} D. Han and D. Larson, Frames, bases and group representations,    {\it Memoirs  Amer. Math. Soc., }  697(2000).

\bibitem{Han-Larson-BLMS} D. Han and D. Larson, Frame duality properties for projective unitary representations, {\it Bull. London Math. Soc.,}   40(2008), 685--695.





\bibitem{HM} T.  Heinosaari, T.  Miyadera, Incompatibility of quantum channels, {\it Journal of Physics A: Mathematical and Theoretical,} 50(2017),  135302.


\bibitem{Jami} A.  Jamiołkowski, Linear transformations which preserve trace and positive semi-definiteness of operator, {\it Rep. Math. Phys.,}  3(4), 275–278 (1972)

\bibitem{KaiLiu-1} K, Liu C. Cheng and D. Han, Orbit injective covariant quantum channels, {\it Lin. Alg. Appl.,}, 2023

\bibitem{KaiLiu-2} K, Liu and D. Han, Phase retrievability of  frames and quantum channels, preprint 2023



\bibitem{MSD-JMP}  M.  Mozrzymas,  M. Studzin'ski and N.  Datta, Structure of irreducibly covariant quantum channels for finite groups,
{\it J. Math. Phys.,}  58, 052204 (2017)

\bibitem{Naimark} M. A. Naimark and A. I. Stern, A. I., Theory of Group Representations, Springer-Verlag, New York, 1982.

\bibitem{Nuw}  M.  Nuwairan, SU(2)-Irreducibly Covariant Quantum Channels and Some
Applications, Ph. D dissertation, University of Ottawa, 2015


\bibitem{Ouyang} Y. Ouyang, Channel covariance, twirling, contraction, and some upper bounds on quantum capacity, {\it Quantum Information and Computation,} 14(2014), 917-936.

\bibitem{5Renes} J. Renes,  R. Blume-Kohout, A. J. Scott and C. Caves  Symmetric informationally complete quantum measurements,  {\it Journal of Mathematical Physics,}  45.6 (2004): 2171-2180.




\bibitem{6Tumalka} R. Tumulka,  POVM (Positive Operator Value Measure), {\it Compendium of Quantum Physics,} (2009): 480-484.

\bibitem{VW}  K. Vollbrecht and R. Werner,
Entanglement measures under symmetry,
{\it Phys. Rev. A,} 64 (6) (2001), Article 062307.



 \end{thebibliography}
\end{document}